\documentclass[twocolumn,runningheads,natbib]{svjour2}
\smartqed  
\usepackage{graphicx}
\journalname{Astrophysics and Space Science}
\begin{document}

\title{Persistent and Transient Blank Field Sources
}
\author{A. Treves\and S. Campana\and M. Chieregato\and A. Moretti\and T. Nelson\and M. Orio}
\institute{A. Treves \at
              Insubria University, via Valleggio 11, Como, Italy \\
              Tel.: +39 031 238 6214\\
              \email{aldo.treves@uninsubria.it}            
           \and
           S. Campana \at
              INAF - Osservatorio astronomico di Brera, Merate, Italy
	   \and
	   M. Chieregato \at
	      Insubria University, Italy -- Z\"urich University, Switzerland
	      	   \and
	   A. Moretti \at
              INAF - Osservatorio astronomico di Brera, Merate, Italy
	      \and
	   T. Nelson \at 
	   University of Wisconsin, USA
	   \and
	   M. Orio \at 
	   INAF - Osservatorio astrofisico di Padova, Italy --  University of Wisconsin, USA	   
}
\date{Received: date / Accepted: date}

\maketitle

\begin{abstract}
Blank field sources (BFS)  are good candidates for hosting dim isolated
neutron stars (DINS). The results of a search of BFS in the ROSAT HRI images
are revised. We then focus on transient BFS, arguing that they belong
to a rather large population. The perspectives of future research on DINS
are then discussed.
\keywords{Stars: neutron}
\PACS{97.60.Jd}
\end{abstract}

\section{Introduction}
\label{intro}
Isolated neutron stars, which have overcome the pulsar phase are elusive
sources. In principle they can shine from some residual internal energy (coolers),
or because of their interaction with the interstellar medium (e.g. accretors).
Their number in the Galaxy should be very high, about one
percent of the total number of stars. Their
significance as a population is of the utmost interest: they are the end-point 
of the evolution of a vast class of stars.

It is just because of these considerations that the discovery of 
dim isolated neutron stars by the ROSAT satellite has been a major 
achievement (see  e.g. Treves et al. 2000, Haberl 2005, Zane et al. 2005).

DINS are one of the main attractions of this meeting, see in particular 
the presentations by Cropper and Popov. Their properties can be summarized
as follows: softness $T\sim100\,$eV; closeness $d\sim100\,$pc; extremely dim optical
counterparts ($V>25$), periodicities of $5-10\,$s; absorption features
below 1 keV. The seven DINS discovered thus far are probably a mixed bag, 
in the sense that the above 
properties may not appear all together. For instance the prototype
of the class 1856--37, has a perfect black body spectrum in the $X$-ray band
with neither absorption lines  nor indications of pulsations. Most likely
they are all coolers (Neuh\"auser \& Tr\"umper 1999, Popov et al.~2002).

In order to further improve our knowledge of DINS it is mandatory to 
enlarge their sample. The procedure followed up to now to discover new DINS
has been to search the ROSAT images for the so called ``Blank Field Sources''
(BFS, Cagnoni et al. 2002), i.e. $X$-ray sources without counterparts 
in other spectral bands, and
 use the properties listed above to argue that the candidate belongs to
the class. This line was pursued by a number of authors, we mention in
particular the recent paper by Agueros et al. (2006) where the ROSAT PSPC
images (RASS) are compared with the Sloan Digital Sky Survey. 

Here we focus on progresses of our search of BFS in
the ROSAT HRI images (Chieregato et al. 2005), concentrating on the 
possible detection of transient BFS.

\section{The ROSAT-HRI Blank Field Sources}
\label{sec:1} 
The ROSAT HRI fields cover $\sim3$\% of the sky but the advantage with respect to
the PSPC is that the position of the source is much better determined,
therefore the limit set by the absence of counterparts can be brought to 
a deeper level.

The $\sim30000$ sources of the ROSAT HRI Brera wavelet catalogue (Panzera et al. 2003)
have been searched for objects {\it a}) with extreme
$f_{\textrm{X}}/f_{\textrm{opt}}$, {\it b}) not too faint, {\it c}) with total
number of photon above a given threshold. Excluding known sources, three
objects have been found which have a statistical significance $>4\sigma$, and
with the closest  counterpart at $>4\sigma$ (see Table \ref{tab:1}). With
respect to Chieregato et al. (2005) 0433+15 was excluded since it was
recognized as spurious. 

\begin{table}[t]
\caption{Blank Field Sources from the HRI Rosat catalogue, adapted from
Chieregato et al.~2005.} 
\centering
\label{tab:1}       
\begin{tabular}{cccccc}
\hline\noalign{\smallskip}
Source&Flux&Prob.&Cts&$f_{\textrm{X}}/f_{\textrm{opt}}$&Opt.\\
&$10^{-13}\frac{\textrm{erg}}{\textrm{cm}^{2}\textrm{s}}$&$\sigma$&&&$\sigma$\\[3pt]
\tableheadseprule\noalign{\smallskip}
$0421-51$&6.5&14.0&742&$>141$&5.3\\
$1357+18$&3.5&4.2 &112&$>47$ &6.3\\
$2007-48$&3.0&4.3 &55 &$>65$ &5.2\\ 
\noalign{\smallskip}\hline
\end{tabular}
\end{table}

The brightest source is 0421--57. It is close to a bright star (see Chieregato
et al. 2005,  Fig.~2). It has been detected with the PSPC at essentially
the same level revealed by the HRI. The source is soft 
($\sim$0.2 keV, see also Section 3) but not as soft as other typical DINS. The two other sources 
are much weaker, and they have not been detected with HRI or PSPC when observed
at different epochs. We will refer to them as transient BFS. Their
light curve have been examined and we can exclude spike-like emission of
duration of seconds or minutes.

\section{New Observations}
\label{sec:2}
A program of $X$-ray observation of the three sources with the Swift XRT is
ongoing (P.I. Moretti). We observed all three BFS: 
0421--57 (11 ks), 1357+18 (9 ks) and 2007--48 (8 ks).
The last two sources were not detected, resulting in 0.3--10 keV $3\,\sigma$
upper limits of $1.8\times 10^{-3}$ counts s$^{-1}$ and $2.2\times 10^{-3}$
counts s$^{-1}$, respectively. Assuming a  power law spectrum with photon
index 2 and a Galactic column density (2.1 and $5.1\times 10^{20}$ cm$^{-2}$)
we obtain $5\times 10^{-14}$ and $8\times10^{-14}$ erg cm$^{-2}$ s$^{-1}$ as
upper limit on the unabsorbed flux in the 0.3--10 keV band.
In the case of 0421--57 we detect the source at a rate of $(2.2\pm0.2)\times
10^{-2}$ counts s$^{-1}$ (about 200 counts) but we are evaluating the
contamination from a bright star closeby. The spectrum is very soft and 
consistent either with a black body ($250\pm30$ eV) or a double Raymond-Smith
model.

Several optical campaigns are now in progress. \linebreak 1357+18 was observed
with the $I$ and $R$ filters and the MiniMo camera with the 3.5 WIYN telescope
in 10 minutes exposures on 2004 June (P.I. M.~Orio). The seeing was about 1.2
arcsec. No optical counterparts were observed in the $3\sigma$ spatial error
circle, 
with a $5\sigma$ upper limit $R>23.4$. VLT ESO observations of 2007--48 were performed in May 2006
(P.I. R.~Mignani), but are not yet analyzed.

\section{Discussion}
\label{sec:3}
We consider in particular the two transient BFS \linebreak (1357+18, 2007--48). Note
that the statistical significance is formally $4\sigma$, and the total number
of photons $\sim100$. It is obvious that one must be extremely cautious
about the reality of the sources, and because they are supposedly transient,
one can't test with further observations.

In the following we  suppose that the sources are real: the optical
counterparts are dim indeed. What can transient BFS be? An extragalactic origin
seems unlikely, because if they were some kind of BL Lac object, a persistent
radio-emission would be expected. Gamma ray bursts are probably to be
excluded too, because as noted above the light curves do not show short term variability. One should
exclude also binaries with a non collapsed companion, because this
should show up in the optical band. One is left to systems consisting of 
collapsed objects. The key point is that the population of which we have
tentatively detected two members could be quite numerous. In fact the
HRI field is $\sim0.2\,$deg$^{2}$, the total exposure time was $\sim3\cdot10^{7}\,$sec. If
the distribution of sources were isotropic this would translate in a rate of 
$10^{5}$ transient BFS per year, otherwise the number should be scaled with the
solid angle. The large parent population points to isolated neutron
stars or white dwarfs. In particular one may wonder if transient BFS are
related to a sudden release of the internal energy of a neutron star, a 
process which may be at work in the recently discovered transient
radio pulsars (McLaughlin et al. 2006; Lyne, this conference).

\section{Conclusions}
\label{sec:4}

In the fifteen years of research about DINS the progress was remarkable,
yet there are two basic points that have not yet been achieved: 

\begin{itemize}
\item There is still no example of a DINS which is convincingly powered
by the accretion of the interstellar medium. While there are a number 
of arguments indicating that these objects should be much rarer
than originally estimated (Treves \& Colpi 1991, Blaes \& Madau 1993,
Perna et  al.~2003 and references therein),
these objects should finally show up, and their emission should be
largely independent of the neutron star age. Neutron stars as old as the
Galaxy could be a part of the lot.  
\item We have not yet
any information on the cousins of DINS, i.e. isolated black holes
(see e.g. Agol \& \linebreak Kamionkowski 2002).
\end{itemize}

The challenge for the future is obviously to explore the sky for
DINS and their cousins, at a flux threshold which is an order of magnitude
lower than that of ROSAT, and which is easily accessible to present
generation $X$-ray telescopes. There is no doubt that the activity up to now has been rather
slow , since rich $X$-ray and optical archives are already available.

 Let us
summarize the hopes for the future, which can derive from a thorough study
of the existing data: 

\begin{itemize}
\item discovery of accretion fed DINS;
\item establishing or excluding the existence of transient BFS; 
\item discovery
of isolated black holes; 
\item discovery of intermediate mass black
holes possibly related to ultraluminous X-ray sources (e.g. Mapelli, Ferrara \& Rea 2006).
\end{itemize}




\end{document}